\documentstyle[preprint,aps]{revtex}

\begin{document}
\bibliographystyle{prsty}
\draft

\title{\bf Quantum computing and communication with decoherence-free atomic states}
\author{Mang Feng 
\thanks{Present address: Institute for Scientific Interchange Foundation, Villa Gualino, Viale Settimio 
Severo 65, I-10131, Torino, Italy. Electronic address: feng@isiosf.isi.it}}
\address{Laboratory of Magnetic Resonance and Atomic and Molecular Physics,\\ 
Wuhan Institute of Physics and Mathematics, Academia Sinica,
Wuhan, 430071, P.R.China}
\date{\today}
\maketitle

\begin{abstract}

To resist decoherence from destroying the phase factor of qubit state, it is important to use
decoherence-free states for processing, transmitting and storing quantum information in quantum 
computing and quantum communication. We propose a practical scheme using four atoms with decoherence-free 
states in a single-mode cavity to realize the entanglement and fundamental quantum logic gates. The 
transmission of quantum information can be made directly from one atom to another, in which the cavity is 
only virtually excited. The possible application and the experimental requirement of our proposal are 
discussed.
\end{abstract}



Both quantum computing and quantum communication have been drawn much attention over past few years. 
It is predicted that quantum computing would outperform the computation made by existing computers in treating some problems, such as 
solving classically intractable problems [1,2] and finding tractable solutions more rapidly [3]. There are many schemes for realizing 
quantum computing with nuclear magnetic resonance (NMR) technique [4], trapped ions [5,6], cavity quantum electrodynamics (QED) [7],
and so on. It has been proven [8] that a suitable combination of some single-qubit rotations and two-qubit operations 
will in principle construct any quantum computing operation we desired. Up to now, some demonstrative
experiments have been made to show the possibility of quantum computing in NMR [9], ion trap [10,11] and cavity-atom system [12].

Besides the application in quantum computing, cavity QED technique is more suitable for quantum communication [13,14]. To our knowledge,
there are some proposals for  teleportation by means of conditional dynamics in cavity-atom systems [15]. Both for the microwave 
cavity and optical one, as long as the cavity with high-quality factor is initially cooled down to the vacuum state, quantum 
information will be transferred from one atom to the cavity , and then from the cavity to another atom. Sometimes the cavity decay 
is also useful for teleportation of an atomic state [16]. Recently, a novel scheme [17] was proposed to use cavities in thermal 
states for teleportation of an atomic state, in which two identical atoms are required to be put into the cavity simultaneously.  

Quantum state is too fragile to exist for a long time unless the system under consideration is perfectly isolated from its environment. 
To overcome the detrimental
effect of decoherence, some methods have been proposed for correcting errors of quantum states after mistakes have been
detected [18], or employing the suitable dynamical coupling to average out the detrimental effect from the environment [19]. 
Alternatively, we can also use some suitable encoded states to resist the decoherence effect [20]. As far as
we know, some studies of logic gates on 
qubits encoded with decoherence-free states have been made before in the context of cavity QED [21] and solid state system [22]. In recent 
experiments with photons [23] and trapped ions [24], the decoherence-free states, which prevent decoherence from destroying the phase 
factors in quantum states of qubits, have been demonstrated to be robust against the collective dephasing. The main idea of these experiments 
is to encode the two-particle quantum state to be $(|eg>+e^{i\theta}|ge>)/\sqrt{2}$ with $|e>$ and $|g>$ being excited and ground 
states respectively of each particle, and $\theta$ the phase factor, so that no phase change would happen under collective dephasing. 
In fact, this idea was also used in a recent work [25] for Grover search with pairs of
trapped ions, in which such a decoherence-free state can also eliminate the change of phase factor due to the time evolution of 
different components of quantum superposition.  

In this contribution, we would demonstrate the implementation of entanglement and fundamental quantum logic
gates with four-particle decoherence-free states 
in the cavity-atom system, if four identical atoms can be put into a single-mode cavity simultaneously. We would show that the 
entanglement of four atoms can be made in one step, much more quickly than any other methods proposed previously. 
More importantly, the realization of fundamental quantum logic gates would open a new way to quantum computing and quantum 
communication with decoherence-free state in the context of cavity QED. Finally we will discuss the application of our scheme by 
taking an example of teleportation.

Consider four identical two-level atoms experiencing simultaneously a monochromatic cavity field, with the Hamiltonian   
\begin{equation}
H = H_{0}+H_{int}
\end{equation}
where $H_{0} = \omega_{a}\sum_{i=1}^{4}\sigma_{iz} + \omega a^{+}a $
and $H_{int} = G\sum_{i\neq j;i,j=1}^{4} (a^{2}\sigma_{i}^{+}\sigma_{j}^{+}+a^{+2}\sigma_{i}\sigma_{j})$
with $\omega_{a}$ and $\omega$ being respectively resonant transition frequency of two levels of each atom and the frequency of
the cavity field. $a^{+}$ and $a$ are creation and annihilation operators of the cavity mode respectively. $G$ is the coupling strength
between the cavity and the atom, and $\sigma_{i}$ $(i=+, -$ and $z)$ are familiar Pauli operators. In the case of large detuning 
between $\omega_{a}$ and $\omega$, there is no energy exchange between atoms and the cavity. So the possible transitions in this 
system are between $|egeg~n>$ and $|gege~n>$, or between $|geeg~n>$ and $|egge~n>$, or between $|eegg~n>$ and $|ggee~n>$ with $|e>$ 
and $|g>$ being excited and ground states of each atom and $|n>$ the cavity state. The Rabi frequency $\Omega$ of these 
transitions, via intermediate state $|gggg~n+2>$ and $|eeee~n-2>$, can be calculated by second order perturbation theory. For example,
for the transition between $|egeg~n>$ and $|gege~n>$, we have
$$\Omega = \frac {<egeg~n|H_{int}|gggg~n+2><gggg~n+2|H_{int}|gege~n>}{\delta} +$$
\begin{equation}
\frac {<egeg~n|H_{int}|eeee~n-2><eeee~n-2|H_{int}|gege~n>}{-\delta}=(4n+2)G^{2}/\delta.
\end{equation}
For other two cases, we can obtain the same value of $\Omega$ as in Eq.(2). Different from the work of two particles [6,17], the Rabi 
frequency in the present system is related to the cavity state. For simplicity, we first assume the cavity state 
to be initially the vacuum state. But this requirement will be loosened, as discussed later. When $n=0$, the effective Hamiltonian of 
Eq.(1) is $H^{eff} = H^{eff}_{0}+H^{eff}_{int}$ with $H^{eff}_{0} =\Omega \sum_{i=1}^{4} |e_{i}><e_{i}|a^{2}a^{+2}$ and 
$$H^{eff}_{int}= \Omega (\sigma_{1}^{+}\sigma_{2}^{+}\sigma_{3}^{-}\sigma_{4}^{-}+
\sigma_{1}^{+}\sigma_{2}^{-}\sigma_{3}^{+}\sigma_{4}^{-} + \sigma_{1}^{+}\sigma_{2}^{-}\sigma_{3}^{-}\sigma_{4}^{+}$$
\begin{equation}
\sigma_{1}^{-}\sigma_{2}^{+}\sigma_{3}^{+}\sigma_{4}^{-}+\sigma_{1}^{-}\sigma_{2}^{+}\sigma_{3}^{-}\sigma_{4}^{+}
+ \sigma_{1}^{-}\sigma_{2}^{-}\sigma_{3}^{+}\sigma_{4}^{+})
\end{equation}
where Pauli operators are written as
$\sigma^{\pm}_{i}=I\otimes\cdots\otimes\sigma^{(i)\pm}\otimes\cdots\otimes I$
with $I=\pmatrix{1 & 0\cr 0 &1}$ and $\sigma^{(i)\pm}$ denoting that the $ith$ matrix is $\sigma^{+}$ or
$\sigma^{-}$. For example,
$\sigma^{+}_{1}=\pmatrix{0 & 1\cr 0 &0}\otimes\pmatrix{1 & 0\cr 0 &1}\otimes\pmatrix{1 & 0\cr 0 &1}\otimes\pmatrix{1 & 0\cr 0 &1}$
and 
$\sigma^{-}_{2}=\pmatrix{1 & 0\cr 0 &1}\otimes\pmatrix{0 & 0\cr 1 &0}\otimes\pmatrix{1 & 0\cr 0
&1}\otimes\pmatrix{1 & 0\cr 0 &1}$.

As $H^{eff}_{0}$ describes Stark shifts related to the photon number, which would become a phase factor in the wave function, in what
following, we would only focus on studying the term of $H^{eff}_{int}$. By means of Eq.(3), the straightforward deduction for the 
time evolution of the system in the interaction representation is 
$$|egeg>\rightarrow\cos(\Omega t)|egeg> -i\sin(\Omega t)|gege>, |gege>\rightarrow\cos(\Omega t)|gege> -i\sin(\Omega t)|egeg>,$$
$$|egge>\rightarrow\cos(\Omega t)|egge> -i\sin(\Omega t)|geeg>, |geeg>\rightarrow\cos(\Omega t)|geeg> -i\sin(\Omega t)|egge>,$$
\begin{equation}
|eegg>\rightarrow\cos(\Omega t)|eegg> -i\sin(\Omega t)|ggee>, |ggee>\rightarrow\cos(\Omega t)|ggee> -i\sin(\Omega t)|eegg>
\end{equation}
which means that we can obtain the four-atom entangled state in one step as long as the four atoms were put into the cavity simultaneously. By
setting different initial conditions of each atoms and controlling the evolution time, we would have different four-atom entangled states.
As the entanglement of four atoms in our scheme can be made much more quickly than any other known methods, our scheme would be useful in
quantum communication for preventing decoherence from affecting quantum channels. 

The four-particle entanglement presented here is different from the entanglement of four trapped ions achieved in NIST [11]. 
The form of the four-ion entangled state is $\Phi_{en}=\frac {1}{\sqrt{2}}(|1111>+|0000>)$ with
$|1>$ and $|0>$ being the excited and ground states respectively of each ion because the trapped ions exchange energy with laser beams 
without changing vibrational states of the ions, and finally get to $\Phi_{en}$. In contrast, in our
scheme, as there is no laser beam in the system, atoms only exchange energy with each other in the case of no change of cavity state. 
Therefore we obtain different forms of four-particle entangled states. Another difference of 
our scheme from Ref.[11] is that, the coupling between atoms and cavity in our
scheme is dependent on the cavity state although the cavity state is only virtually excited in atomic state transitions. In the treatment
above, we supposed the cavity state to be initially vacuum. While for the cavity state with $n=1,2,\cdots$, our scheme could be still
available. However, if $n$ is too large, the perturbation theory will be invalid for solving the coupling $\Omega$. In this sense,
even if the cavity state is in thermal state, our proposal would still work as long as the mean photon number in the cavity is small. 

Similar to Ref.[6] that the controlled-NOT (CNOT) gate has been constructed on two identical ions experiencing a single-mode laser beam, 
we can also carry out the CNOT gate on the four identical atoms based on Eq.(4). We choose four states $|egeg>_{1234}$, $|gege>_{1234}$, 
$|egge>_{1234}$
and $|geeg>_{1234}$ to form the computational space where subscripts mean the labeling of the atoms. Following sequence of
operation: $H_{34}, P_{34}, R, P_{34}, H_{12}, H_{34}$ and $P^{-1}_{34}$ will yield  the CNOT gate as 
$|eg>_{12}|ge>_{34}\rightarrow |eg>_{12}|ge>_{34}$, $|ge>_{12}|ge>_{34}\rightarrow |ge>_{12}|ge>_{34}$, 
$|eg>_{12}|eg>_{34}\rightarrow |ge>_{12}|eg>_{34}$ and
$|ge>_{12}|eg>_{34}\rightarrow |eg>_{12}|eg>_{34}$, where $H_{ij}$ is made by putting the atoms of $i$ and $j$ into a single-mode cavity
simultaneously with the transformations $|eg>_{ij}\rightarrow (|eg>_{ij}-i|ge>_{ij})/\sqrt{2}$ and 
$|ge>_{ij}\rightarrow (|ge>_{ij}-i|eg>_{ij})/\sqrt{2}$ [17]. $P_{ij}$ is a $\frac {\pi}{2}$ phase change of $|eg>_{ij}$ and $R$ is the 
evolution shown in Eqs.(4) with $\Omega t=3\pi/4$.

As the Hadamard transformation on two atoms can be realized by $H_{ij}$ in addition to $P_{ij}$, the realization of CNOT gates means that 
any quantum computing operation can be constructed with decoherence-free states in cavity-atom systems. Meanwhile, our scheme would also
have important application in teleportation. In the original proposal of
teleportation, an unknown quantum state could be perfectly sent from Alice to Bob over arbitrary distance by a measurement of Alice and 
suitable operations of
Bob, in which an entangled state shared previously by Alice and Bob and two bits of classical message sent from Alice to Bob are necessary.
However, if we want to teleport an atomic superposition state with components of different eigenenergies, the receiver would 
probably obtain another state instead of the desired one, because of the time evolution of component states. For example, Alice wants 
to send the state of $\psi=|g>+e^{i\theta}|e>$ to Bob, where $|g>$ and $|e>$ are different eigenstates respectively. According to 
Schr\"odinger equation, during a time interval $t$, $\psi$ will evolve to $e^{-iE_{g}t/\hbar}|g>+ e^{i\theta}e^{-iE_{e}t/\hbar}|e>$ with 
$E_{g}$ and $E_{e}$ being the eigenenergies of states $|g>$ and $|e>$ respectively. So if Alice takes a period of T to inform Bob 
about her measurement 
result, Bob will obtain $|g>+ e^{i\theta}e^{-i(E_{e}-E_{g})T/\hbar}|e>$. It means that Bob could not obtain the desired state $\psi$
unless teleportation is made superluminally (i.e., $T=0$) or in the system with $E_{g}=E_{e}$. This evolution problem of eigenstates 
exists in all teleportation schemes with cavity QED [15-17]. But no consideration has yet been paid on this respect. While if teleported 
state is set to be $\psi=|ge>+e^{i\theta}|eg>$, and Bell states are $\Phi^{\pm}=(|egeg>\pm i|gege>)/\sqrt{2}$ and 
$\Psi^{\pm}=(|egge>\pm i|geeg>)/\sqrt{2}$, by supposing $|eg>=|\tilde{1}>$ and $|ge>=|\tilde{0}>$, we can perfectly realize the
teleportation idea in Ref.[13] with cavity QED technique. We should emphasize that Bell states can be fully distinguished in our
scheme, instead of only $\Psi^{\pm}$ being identified as in Ref.[17]. If we put the four atoms with Bell states into a single-mode cavity for a
period of $t=\pi/4\Omega$, we have the time evolution yielding $\Phi^{+}\rightarrow |egeg>$, $\Phi^{-}\rightarrow -i|gege>$, 
$\Psi^{+}\rightarrow |egge>$ and $\Psi^{-}\rightarrow -i|geeg>$. So the individual measurement on the atoms would make us identify each
of Bell states.  

Let us discuss the possible error happening in the implementation of the CNOT gate. Our scheme for CNOT gate includes both two-atom and 
four-atom operations. Any inadequate operation on two-atom pairs (with $H_{ij}$ and $P_{ij}$) or four atoms (with $R$) would
produce errors in the CNOT gate. For convenience of our discussion, we consider a simple case. Suppose when we put four atoms with
$|egeg>_{1234}$ into the cavity, an imperfect manipulation takes place, i.e., two atoms 1 and 2 being put into the cavity
a little sooner than the atoms 3 and 4, which yields
$\Psi=\cos (\lambda t_{1})\{\cos [\Omega (t-t_{1})]|egeg>_{1234} -i\sin [\Omega (t-t_{1})]|gege>_{1234}\}
-i\sin (\lambda t_{1})\{\cos [\Omega (t-t_{1})]|geeg>_{1234} -i\sin [\Omega (t-t_{1})]|egge>_{1234}\}$
where $\lambda=\Omega/2$ [17], $t_{1}$ is the time for atoms 1 and 2 in the cavity before the 
atoms 3 and 4 are input, and we neglect the common phase factor. If we hope to have a operation $R$, we will find that the fidelity is
$<\Psi_{R}|\Psi>\approx 0.98$ in the case of $t_{1}=0.02 t$ with $\Psi_{R}$ the desired state under the operation $R$. So our
proposal still works very well [26]. 
Moreover, how to distinguish the four identical atoms is essential to our scheme. Above discussion has shown that any confusion in the
manipulation of atoms will make our proposal incorrect or even invalid. Fortunately, four identical atoms in a cavity are distinguishable 
with the present cavity QED technique, because the atoms in the cavity can be manipulated at centimeter-scale distance [27].  
Furthermore, the spontaneous emission of the excited state is another possible detrimental effect in
our scheme. From the discussion above, we know that the time for achieving a maximal entanglement and a CNOT gate of four atoms are
respectively $\pi\delta/8G^{2}$ and $7\pi\delta/8G^{2}$. As $G=2\pi\times 47 kHz$ and the lifetime of the excited state of a Rydberg atom
is generally $3\times 10^{-2} s$ [27], if $\delta=10 G$, the time of above implementations would be on the order of $10^{-4} s$, which is
much shorter than the lifetime of the excited state of a Rydberg atom. So our scheme is practical.

To summarize, a practical proposal for quantum communication and fundamental gates of quantum computing with four-atom decoherence-free 
states has been put forward, based on the supposition that we can simultaneously put four identical atoms into a single-mode cavity 
with few photons. 
As the entanglement and quantum gates are made between atoms, instead of between atoms and photons, the cavity state is only virtually
excited so that the cavity is not strictly required to be initially in vacuum state and the quality factor of the cavity is not 
necessarily set to be very high [17]. Moreover, the information stored in atoms with decoherence-free 
states is more safe than that stored in photons, and the implementation on atoms for processing, transmitting or storing quantum 
information is easier than that on photons. Furthermore, the former works involving photons had no way to completely measure Bell
states due to the lack of photon-photon interaction [28]. While in our scheme, the effective interaction in
the cavity-atom system guarantees the 
complete measurement of Bell states. Besides although the four-particle entanglement has been realized previously in the ion
trap, the four-particle entanglement proposed here is different and is more important because it corresponds to decoherence-free
states. While except the dephasing, we have not
considered any other effects of decoherence in our scheme, because the collective dephasing is much stronger than other effects of
decoherence [23,24] when the atoms are closed enough. Compared with previous proposals [7,17], our scheme is not more stringent. 
It can be experimentally achieved with the cavity QED technique presently or in the near future. We hope that  our proposal
would be useful for perfect information processing, transmission and storage in quantum computing and quantum communication. 

Valuable discussion with Jianwei Pan is highly acknowledged.
The author also would like to thank the hospitality of Max Planck institute for the Physics 
of Complex Systems, where part of this work was carried out. The work is partly supported by the National Natural Science Foundation of China.


\begin{thebibliography}{27}
\bibitem{26} For other cases, such as one atom is input before other three and so on, the calculation is a little complicated.
But as long as the time $t_{1}$ is small enough, for example, less than $0.05t$, the fidelity would be still very high.
\end{thebibliography}
\end{document}